\documentclass[sigconf]{acmart}
\usepackage{enumitem}
\usepackage{subcaption}

\usepackage{amsmath, amssymb, amsthm}
\usepackage{longtable}
\usepackage{multirow}
\usepackage{listings}
\usepackage{verbatim}
\newcommand{\eat}[1]{}

\newtheorem{definition}{Definition}
\AtBeginDocument{%
  }

\copyrightyear{2026}
\acmYear{2026}
\setcopyright{cc}
\setcctype{by}
\acmConference[SeQureDB '26]{Workshop on Secure and Private Data Management}{May 31-June 05, 2026}{Bengaluru, India}
\acmBooktitle{Workshop on Secure and Private Data Management (SeQureDB '26), May 31-June 05, 2026, Bengaluru, India}
\acmDOI{10.1145/3807894.3810277}
\acmISBN{979-8-4007-2219-6/2026/05}

\begin{document}

\title{Compliance in Databases: A Study of Structural Policies and Query Optimization}

\author{%
  Ahana Pradhan\textsuperscript{*},\;
  Srinivas Karthik\textsuperscript{**},\;
  Imtiyazuddin Shaik\textsuperscript{*},\;
  Srinivas Vivek\textsuperscript{*}%
}

\affiliation{%
  \textsuperscript{*}IIIT Bangalore, India;\;
  \textsuperscript{**}Microsoft
    \country{India} 
}

\email{%
  ahana.pradhan@iiitb.ac.in, srv@microsoft.com, shaik.imtiyazuddin@iiitb.ac.in, srinivas.vivek@iiitb.ac.in%
}

\renewcommand{\shortauthors}{A. Pradhan, S. Karthik, I. Shaik, S. Vivek}

\begin{abstract}
\eat{
Modern privacy regulations require fine-grained, context-sensitive control over data access, making database enforcement of compliance policies a systems problem. Existing database mechanisms, including row-level security, views, and query rewriting, can express simple content-based access control rules, but they do not adequately support the structural complexity of real compliance requirements or the composition of multiple regulatory policies. In practice, policies range from local filters to joins, anti-joins, correlated subqueries, aggregates, and threshold-based predicates, each affecting query optimization differently. This paper argues for a systematic policy-structure formulation framework and introduces a taxonomy of atomic SQL compliance policies grounded in query-graph structure. We further present a grammar-based approach for composing multiple policies into analyzable policy trees and evaluate how a database engine, PostgreSQL, handles such composed policies under enforcement mechanisms. Our study exposes the performance sensitivity of compliance enforcement to policy structure and implementation choice.
}
Growing privacy regulations and internal governance mandates are driving demand for fine-grained, context-sensitive access control in data management systems. Among competing approaches, content-based access control—where access decisions depend on the data values referenced by a query—is becoming particularly prominent, and is supported directly in modern database engines. 

While simple content-based predicates often incur negligible overhead, increasingly rich policies can interact in subtle ways with query optimization, leading to significant and poorly understood performance variability. This paper investigates this gap by introducing a structural framework and expressive policy grammar for modeling content-based compliance policies and analyzing their impact on query planning and execution in database systems.

Building on this framework, we augment an analytical benchmark with structured policy workloads, enabling controlled evaluation of enforcement mechanisms and optimization strategies under combined query–policy workloads. Our experimental results show that policy structure has a decisive impact on optimizer behavior and end-to-end performance, underscoring the need for policy-aware database and optimizer design.

\end{abstract}

\ccsdesc[500]{Security and privacy~Information accountability and usage control}

\maketitle

\section{Introduction}
\label{sec:introduction}

Data privacy has become a first-class concern in modern database systems. Regulations such as GDPR, HIPAA, and CCPA impose strict requirements on how organizations store, access, and process personal and sensitive data \cite{HoofnagleVanDerSlootZuiderveenBorgesius2019}. Beyond regulatory compliance, internal governance policies increasingly demand that access to data be conditioned not merely on a user's role or identity, but on the \emph{content} of the data itself. For instance, a supplier record may be visible only if the supplier operates in an approved region; an order may be accessible only if it carries sufficient priority. This model is known as \emph{content-based}  access control ({aka} \emph{fine-grained} access control) ~\cite{RizviMendelzonSudarshanRoy2004,pappachan2020sieve, poepsellemaitre2024disclosure}. Here the access decisions are a function of the data values in a query, rather than a static privilege table.

Content-based access control is increasingly supported as a native primitive in relational database engines. PostgreSQL implements it through Row-Level Security (RLS)~\cite{PostgreSQLRLS2025}, which allows administrators to attach per-table filter predicates that are automatically injected into every query accessing that table. Similar mechanisms exist in SQL Server~\cite{rls1}, Oracle (VPD)~\cite{OracleVPDGuide}, Snowflake~\cite{snowflake_rls}, and other commercial systems. The appeal of native enforcement is clear: policies are defined once, applied transparently, and cannot be bypassed by application code.

\subsubsection*{\textbf{The problem.}}
Despite widespread adoption, the performance implications of content-based policies remain poorly understood. Simple policies —a single equality predicate or a range condition - are typically absorbed by the optimizer without measurable overhead. However, realistic governance scenarios require \emph{compositional} policies: a policy on \texttt{lineitem} may depend on a policy on \texttt{partsupp}, which in turn depends on a policy on \texttt{supplier}. As such tiered, cross-table dependencies accumulate, the injected predicates grow in structural complexity. Standard cost-based optimizers produce policy-violating plans in the majority of cases when compliance constraints are present~\cite{beedkar2021compliant}, underscoring that policy-unawareness is a fundamental optimizer design gap. When enforcement is added, the optimizer must simultaneously reason over the original query and injected policy subqueries—a task current planners were not designed for. The problem is further compounded by the diversity of enforcement mechanisms available: native RLS, security barrier views, query rewriting, and user-defined functions each impose different boundaries on what the optimizer can see and transform, making different trade-offs between security guarantees, optimization transparency, and implementation complexity.

\subsubsection*{\textbf{The gap.}}
The absence of a structured framework for generating and reasoning about policy complexity means that prior work either evaluates enforcement strategies on trivially simple policies that do not stress the optimizer, or on ad-hoc application-specific policies that resist generalization. There is no established benchmark or policy generation framework that allows controlled, reproducible evaluation of enforcement mechanisms as policy complexity scales.

\subsubsection*{\textbf{Contributions:}}We address this gap through three contributions.

\begin{enumerate}[wide, labelwidth=!, labelindent=0pt]

\item \textbf{A structural policy generation framework.} At first glance, LLMs appear well-suited for policy generation given their NL2SQL capabilities. However, naive LLM-generated policies suffer from limited scope policies and cyclic/self-referencing policies, which are unusable through the basic enforcement methods.

Motivated by this, we introduce a structural model and expressive grammar for content-based compliance policies to capture a wide range of practical access control scenarios. The framework explicitly supports both \emph{white-box} enforcement—where policies are acyclic and admit cross query-policy optimization—and \emph{black-box} enforcement—where cyclic, unrestricted policies are safely handled by treating policy predicates as opaque to the optimizer.

\item \textbf{A policy-augmented benchmark.} We extend the TPC-H analytical benchmark with structured policy workloads generated using our framework. Policies are assigned business semantics over the TPC-H schema and systematically vary in composition depth, enabling controlled evaluation across a spectrum of complexity. The benchmark has been made publicly available\footnote{https://github.com/imtiyazuddin/secure26db.git}.

\item \textbf{A comparative empirical study.} We evaluate four enforcement strategies available in PostgreSQL—Pure RLS (predicate injection \& User Defined Functions),  Indexed RLS, Secure Views, and Query Rewrite via \texttt{LITHE}~\cite{lithe}—under both acyclic and cyclic policy workloads, measuring execution time, planning overhead, and robustness. Our results show that policy structure primarily determines enforcement cost, that no single strategy dominates across all settings, and that policy-aware indexing and query rewriting offer the most reliable paths to scalable enforcement. \textit{The evaluation lays the groundwork for a new class of policy-aware query optimizers that treat enforcement predicates as first-class optimization objects.}

\end{enumerate}

\section{A Framework for SQL Policy Formulation}
\label{sec:policy-taxonomy}

\begin{figure}
    \centering
    \includegraphics[width=\linewidth]{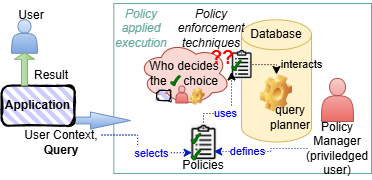}
    \caption{Interplay Between Users, Policies, and Databases}
    \label{fig:big_picture}
\end{figure}

\subsection{Motivation}

Query processing engines require a formal framework that treats regulatory policies as first-class citizens—essential for ensuring legal compliance, protecting privacy, enabling provable correctness, and providing scalable, uniform enforcement. 

The framework's complexity arises from the interplay of multiple stakeholders and competing concerns as captured in Figure~\ref{fig:big_picture}. \emph{Policy managers} define the compliance rules—often independently of the queries that will be subject to them—while \emph{database administrators} must translate these rules into enforceable mechanisms within the engine. \emph{End users} and application developers, in turn, expect transparent enforcement without changes to their query logic. This separation of concerns creates a rich design space: the same policy can be enforced in multiple ways, each with different implications for security, performance, and implementation effort.

From a \emph{security} standpoint, the choice of mechanism determines how much of the policy is visible to the optimizer. Black-box mechanisms—such as \texttt{SECURITY DEFINER} UDFs in PosgreSQL —fully hide policy logic from the planner, preventing information leakage through plan structure but forgoing cross query-policy optimization. White-box mechanisms—such as RLS and query rewriting—expose policy predicates to the optimizer, enabling better plans at the cost of increased attack surface for inference.

From a \emph{performance} standpoint, policy persistence matters. Stable, long-lived policies justify the upfront cost of materialized views, policy-coverage indexes, or precompiled UDFs. Ad-hoc or frequently changing policies are better served by query rewriting, which incurs no materialization cost but demands a capable rewriter. Policy composition depth further amplifies this tension: tiered, multi-table policies generate deeply nested subqueries that stress cardinality estimation and join ordering regardless of the enforcement mechanism. 

This interplay of stakeholder roles, security requirements, performance trade-offs, and enforcement diversity motivates the need for a principled policy framework. Current engines lack the abstractions to reason jointly about policy structure and query optimization—policies are either too shallow to reflect real governance requirements or too complex for the optimizer to handle efficiently. A formal framework spanning atomic to composite policies and white-box to black-box enforcement is therefore essential for understanding and closing this gap, which is described next.

\subsection{Policy Model}\label{sect:atomic}
 Let $\mathcal{D} = (R_1, R_2, \ldots, R_n)$ be a relational database schema, where each relation $R_i$ has attributes $\text{attr}(R_i)$ = $\{A_1$, $A_2,$ $ \ldots, A_{k_i}\}$. Let $I$ denote a database instance over $\mathcal{D}$. Let $\mathcal{U}$ denote the set of users (subjects) and $\mathcal{C}$ denote the execution context (including user identity, role, session attributes, and environment variables).

\paragraph{Semantics of ``Policy" w.r.t. user query} In this paper, we restrict ourselves to policy logic expressible in SQL. Under this assumption, an {\em enforced-policy} can be viewed as a {\tt SELECT} query: the {\tt WHERE} clause (the $\sigma$ operator) governs tuple visibility, while the {\tt SELECT} clause (the $\pi$ operator) governs attribute presentation, i.e., what the user can observe when the policy is evaluated. When multiple such policies are defined over a table, they can be combined through different logical compositions. 

A policy $\alpha := \langle R, Q, \sigma, \pi, \mu \rangle$ represents a basic policy unit {\em on relation schema $R$} to restrict its data to the user who queries it. The semantics of $\langle R, Q, \sigma, \pi, \mu \rangle$ is a relational expression, which is explained using the following three definitions.  

\begin{definition}
[Masking Function]\label{def:mu} A masking function $\mu: \mathrm{dom}(A_j) \to \mathrm{dom}(A_j) \cup \{\bot\}$ 
maps each value of attribute $A_j$ of $R$ either to a (possibly transformed) 
value in the same domain or to $\bot$, which denotes 
a suppressed value.  
\end{definition}  

\begin{definition}
[Tuple Visibility] A tuple $t \in R_i^I$ is \emph{visible} to a user $u \in \mathcal{U}$ in context $\mathcal{C}$ if the enforcement mechanism does not suppress it from query results. 
\end{definition}

Therefore, a tuple $t$ $\in$ $R^I_i$ is visible under policy $\alpha$ if masking function $\mu$ does not map all attribute values of $t$ to $\bot$.  

\begin{definition}
[Attribute Presentation] For a visible tuple $t$, an attribute $A_j$ is \emph{presented faithfully} if its original value in the database is returned to the user; it is \emph{transformed} if a masking function $\mu(t.A_j)$ is applied (e.g., redaction, generalization, anonymization). 
\end{definition}

\begin{definition}[Atomic Policy]
Let $R$ be a relation schema and let $Q$ be a predicate over $R$.
An \emph{atomic policy}
$
\alpha = \langle R, Q, \sigma, \pi, \mu \rangle
$
where:
\begin{itemize}[wide, labelwidth=!, labelindent=0pt]
\item $\sigma$ is the standard  selection operator over $R$ with predicate $Q$,
    \item $\pi$ is a standard projection operator over schema $R$,
    \item $\mu$ is a masking projection operator over schema $R$, which, for each attribute $A_i \in R$, it applies a masking function 
$f_i$ (Definition~\ref{def:mu}).
\end{itemize}

The semantics of $\alpha$ over a database instance $D$ is defined as:
\[
\llbracket \alpha \rrbracket_D
=
\pi\big(\sigma_Q(R^D)\big)
\;\cup\;
\mu\big(\sigma_{\neg Q}(R^D)\big).
\]
\end{definition}   

\subsubsection*{Example set of atomic policies}
Structurally, policy predicate query $Q$ can differ based on its SQL constructs. In this paper, we specify the following kinds of atomic policy structures to illustrate the working principle of the framework.

\begin{enumerate}[wide, labelwidth=!, labelindent=0pt]
    
\item {\bf Attribute Predicate}.
The class of queries constrain visibility using attribute value thresholds. Such policies arise naturally in geographic,
organizational, and role-based compliance rules. One or more relation can be included in a policy, whose attribute values can be predicated.

\item {\bf Existential Predicate}.
Existence-based policies grant visibility if at least one related tuple satisfies
a qualifying predicate. These policies correspond to \emph{semi-join semantics}.

Note that attribute predicates -- category (1) -- can be expressed using EXISTS in SQL, as shown in the corresponding example in Appendix; they semantically enforce attribute constraints across joined relations, whereas existential predicates require only the existence of a qualifying related tuple.

\item {\bf Universal Predicate}.
Universal policy predicates enforce violation-free constraints: an entity is visible only if
no related tuple violates the policy. These are typically expressed using
\texttt{NOT EXISTS} and correspond to \emph{anti-join semantics}.

\item {\bf Grouping/Aggregated Predicate}.
This class of policies enforces compliance based on statistics computed over groups of tuples,
such as counts or distinct counts. These policies use \texttt{GROUP BY} and \texttt{HAVING}.

\item {\bf Statistical Predicate}.
Statistical predicated policies constrain higher-order statistics, often represented by nested aggregates, and other statistical functions beyond classical
aggregates, capturing correlation-like relationships indicative of specific behavior.

Example SQLs for each of the above policy structures are listed in the Appendix on Page \pageref{Sect:appendix}. 
\end{enumerate}

\eat{
\subsection{Policy Formulation Framework Layers}
The above classes of atomic policy structures form the layer 1 -- the bottom most -- in our framework. The chosen masking function specifies the appearance of the policy-filtered tuples. Thus, it forms an orthogonal layer wrt layer 1, which we refer to as the {\em Enforcement Action} type. Above this, we have a grammar to compose already defined policies -- either from layer 1 or from layer 2. Importantly, multiple policies are {\em compatible to compose} when their choice of the enforcement policies is the same, even when the actual projection functions are different.
Below is the overall architecture. 
\begin{itemize}
    \item \textbf{Layer 1: Atomic Policies}
    \begin{itemize}
        \item P1: Attribute predicate
        \item P2: Existential predicate
        \item P3: Anti-existential predicate
        \item P4: Aggregate predicate
        \item P5: Statistical predicate
    \end{itemize}

    \item \textbf{Enforcement Action}
    \begin{itemize}
        \item Filter
        \item Transform
    \end{itemize}

    \item \textbf{Layer 2: Composed Policies}
    \begin{itemize}
        \item C1: Conjunctive
        \item C2: Disjunctive
        \item C3: Negated
        \item C4: Conditional
    \end{itemize}

\end{itemize}

Layer 1 and Enforcement Action are already defined above. Now we proceed to describe Layer 2.
}

\subsection{Policy Grammar: Composite Policies}
To construct new policies by combining multiple atomic policies, a {\em policy grammar} is specified by the policy manager. For instance, we use the following grammar which allows logical combination of policies in an expressive way: A {policy} $\mathcal{P}$ is defined recursively by a context-free grammar. $\mathcal{P}$ is either a {relational expression} $\beta$ or a {conditional relational expression} $\iota$.
\begin{itemize}[wide, labelwidth=!, labelindent=0pt]
    \item A relational expression $\beta$ may be:
 an {atomic policy query} $\alpha$,
     {\em a conjunction or a disjunction of the predicate queries} of two atomic policies, or 
     {\em a negation of the predicate query} of an atomic policy.
     \item A conditional expression $\iota$ has the form:
$\textbf{if } Q \textbf{ then } \mathcal{P}_1 \textbf{ else } \mathcal{P}_2$,
where the resulting policy depends on the truth value of the boolean predicate query $Q$.

\end{itemize}

\[
\begin{aligned}
\mathcal{P} &::= \beta \mid \iota \\
\beta &::= \alpha \mid (\beta \wedge \beta) \mid (\beta \vee \beta) \mid (\neg\, \beta) \\
\iota &::= \textbf{if } Q \textbf{ then } \mathcal{P} \textbf{ else } \mathcal{P} \\
\alpha &::= \langle R_i, Q, \sigma, \pi, \mu \rangle
\end{aligned}
\]

To elaborate the composition using logical operators, let us take the instance of {\em AND-ing} two atomic policies $\alpha_1$ = $\langle R_1, Q_1, \sigma, \pi, \mu_1 \rangle$ and $\alpha_2$ = $\langle R_2, Q_2, \sigma, \pi, \mu_2 \rangle$. They are {\em compatible} to compose only when $R_1 = R_2$, i.e., both the policies are on the same relation. Consequently, $\pi$ is the same identity projection for both $\alpha_1$ and $\alpha_2$. The masking functions of $\mu_1$ and $\mu_2$ may be {\em different}. And lastly, predicates $Q_1$ and $Q_2$ {\em are} different to motivate a meaningful composition. 

Therefore, the logical AND-ing has the following semantics:
\[
\llbracket \alpha_1 \wedge \alpha_2 \rrbracket_D
=
\pi\big(\sigma_{Q_1\wedge Q_2}(R^D)\big)
\;\cup\;
\mu_1\big(\sigma_{\neg Q_1}(R^D)\big)
\;\cup\;
\mu_2\big(\sigma_{\neg Q_2}(R^D)\big).
\]

Similarly, the logical OR-ing makes a tuple fully visible when either of the policy predicates -- $Q_1$ or $Q_2$ -- holds. Only when both predicates fail does masking apply, and in that case both masking operators act on the region where the tuple satisfies neither predicate.
\[
\begin{aligned}
\llbracket \alpha_1 \vee \alpha_2 \rrbracket_D
&= \pi\big(\sigma_{Q_1 \vee Q_2}(R^D)\big) \\
&\;\cup\; \mu_1\big(\sigma_{\neg Q_1 \wedge \neg Q_2}(R^D)\big)
\;\cup\; \mu_2\big(\sigma_{\neg Q_2 \wedge \neg Q_1}(R^D)\big).
\end{aligned}
\]

Next, negation swaps the visibility and masking regions: tuples that satisfy $Q$
are masked, while tuples that do not satisfy $Q$ are made fully visible. Formally: 
\[
\llbracket \neg \alpha \rrbracket_D
=
\pi\big(\sigma_{\neg Q}(R^D)\big)
\;\cup\;
\mu\big(\sigma_{Q}(R^D)\big).
\]

Note that the conditional expression can be equivalently written as implication using the other grammar rules. However, it adds {\tt CASE-WHEN-ELSE-THEN} SQL-construct in this grammatical formulation, which eliminates the costly use of negation of a policy query.  
The above structural variations are sufficient to capture real-world analytic queries, as they capture flat SPJGAO, nested subqueries and correlations. Consequently, they are also sufficient to capture meaningful complex business rules.

\subsubsection*{Composite Policy Derivation Example}\label{sect:example}

Let us elaborate on the policy derivation using an example.
\paragraph{Policy Intent.}
The visibility of a customer tuple depends on the customer’s total spending and return behavior. 
A tuple is fully visible if the customer has spent more than 5000 in total and has no store returns. 
It is partially masked if the customer has spent more than 5000 but has at least one store return. 
In all other cases, the tuple is fully masked.

\paragraph{Base Relation.}
$R = \textit{customer}.
$

\paragraph{High-value predicate.}

$Q_1(c)$ := {\footnotesize
\begin{verbatim}
(SELECT COALESCE(SUM(l.l_extendedprice * (1 - l.l_discount)), 0)
  FROM orders o, lineitem l WHERE l.l_orderkey = o.o_orderkey
   AND o.o_custkey = c.c_custkey) > 5000
\end{verbatim}
}

\paragraph{Return predicate.}
$Q_2(c)$ := {\footnotesize
\begin{verbatim}
EXISTS (SELECT 1 FROM orders o, lineitem l
    WHERE l.l_orderkey = o.o_orderkey
    AND o.o_custkey = c.c_custkey AND l.l_returnflag = 'R')
\end{verbatim}
}

\subsubsection*{Atomic Policies. ($\mu_{\emptyset}$ indicates no masking/transformation)}

\[
\alpha_1 = \langle R, Q_1 \wedge \neg Q_2, \sigma, \pi, \mu_{\emptyset} \rangle
\]
\[
\alpha_2 = \langle R, Q_1 \wedge Q_2, \sigma, \pi, \mu_{\text{partial}} \rangle
\] \[
\alpha_3 = \langle R, \neg Q_1, \sigma, \pi, \mu_{\text{NULL}} \rangle
\]

\noindent{Composed Policy}
$
\mathcal{P}
=
\textbf{if } Q_1
\textbf{ then }
\left(
  \textbf{if } Q_2
  \textbf{ then } \alpha_2
  \textbf{ else } \alpha_1
\right)
\textbf{ else } \alpha_3
$

\noindent
$
\llbracket \mathcal{P} \rrbracket_D
=
\pi(\sigma_{Q_1 \wedge \neg Q_2}(R^D))
\;\cup\;
\mu_{\text{partial}}(\sigma_{Q_1 \wedge Q_2}(R^D))
\;\cup\;
\mu_{\text{NULL}}(\sigma_{\neg Q_1}(R^D)).
$

\noindent{The corresponding SQL is listed in the Appendix.}

\paragraph{Takeaway.}
The grammar captures a wide range of practical access control scenarios—from simple attribute filters to aggregate-guarded and statistically conditioned policies—while remaining grounded in standard relational algebra. Every policy admits a canonical AST representation that exposes its syntactic structure to program-transformation techniques, enabling predicate pushdown, redundancy elimination, and joint query–policy optimization.



\section{Policy Enforcement Strategies}\label{sect:pol_imple}

The compliance policies must be enforced inside the database engine to ensure correct behavior under any query. The following mechanisms are available in any standard database engine, which can be used to implement such policies.

\begin{itemize}[wide, labelwidth=!, labelindent=0pt]
\item {\em RLS-based Enforcement.}
Row-level security integrates policy chec\-ks directly into the database engine. Users write queries normally, and the system enforces the restriction transparently by adding the policy predicate to the query under the hood. 
Database engines such as PostgreSQL and SQL Server provide \emph{row security policies} that automatically append policy predicates to query plans\cite{PostgreSQLRLS2025}, \cite{MicrosoftSQLServerRLS}. 
Oracle provides the analogous mechanism through \emph{Virtual Private Database (VPD)}, which dynamically appends security predicates based on session and application context~\cite{OracleVPDGuide}.

\item {\em Query Rewrite-based Enforcement.} While the policy predicate-injected user query is optimized natively by the database engine, a layer of semantic rewrite has the potential to improve the plan quality of the overall query. Consequently, a rewrite of the RLS-produced query is also a potential candidate mechanism.

\item {\em View-based Enforcement.}
Views enforce policies by embedding the compliance logic inside a predefined query. Users access the view instead of the base table, so the policy predicates are automatically applied. This keeps policy logic separate from user-written queries and makes the rules easy to see and audit at the schema level. Policies can also be composed by defining views on top of other views without changing the underlying tables.

\item {\em User-Defined Function-based enforcement.}
UDFs can be used to modularize policy logic by allowing it to be defined once and invoked uniformly across user queries. This promotes reuse and simplifies maintenance: when a regulation changes, only the UDF implementation needs to be updated. In addition, UDFs can express fine-grained restrictions that depend on runtime identity, session state, or other contextual information, thereby enabling dynamic and context-aware enforcement. 
\end{itemize}

\subsection{White-box and Black-box Enforcement}
Enforcement methods fall into two broad categories. In \textit{white-box} enforcement, the policy predicate is visible to the query optimizer, enabling joint query--policy optimization; predicate pushdown, index exploitation, and partition pruning are all applicable. RLS, query rewriting, views, and transparent UDFs belong to this class. In \textit{black-box} enforcement, the policy logic is opaque to the planner—typically for security reasons (side channel or error-based inference attacks) where users lack permission to inspect the policy—disabling cross-policy optimization. This class includes opaque views and engine-specific black-box UDFs.

\eat{

While RLS is the basic enforcement strategy, and hence, easy to use, the engine-defined implementation offers limited flexibility for policy-specific optimization or semantic rewriting. Query rewrite eliminates this limitation, but it introduces additional system complexity and requires strong guarantees of correctness. On the other hand, large numbers of evolving or layered views can become cumbersome to maintain and may degrade optimization effectiveness. Lastly, UDFs are the most usable for policy combinations, but pose optimization barrier thereby limiting performance.

The composite policy discussed in Section~\ref{sect:example} is equivalent to the native RLS, which transparently inlines the policy predicates. For this example, we have additional CTEs that express the policy. 
The {\em enforcement action} type of a policy -- filter or transformation -- the functions returned by $\mu$ -- is the primary factor in necessitating a view-based implementation over an RLS-based one. For instance,  when $\mu$ gives $\bot$ for the tuples that do not satisfy policy predicate $Q$ -- expressing tuple suppression -- both implementations are applicable. However, when $\mu$ is a masking function, like the ones present in the example in Section~\ref{sect:example}, RLS is not applicable since {\em it is a filtering technique by construction}. Therefore, either the view-based or rewrite-based implementations are the options we are left with.
}

\subsection{Note on Policy Structure}
\subsubsection*{Cyclic} A policy predicate may reference the very table to which it is attached
(\emph{direct self-reference}), or it may induce a reference cycle
through a chain of intermediate tables (\emph{multi-hop self-reference},
e.g.\ \(t_1 \to t_2 \to \cdots \to t_1\)).
In either case, the enforcement mechanism re-invokes the policy predicate
while evaluating it, typically manifesting as an \emph{infinite
recursion} error during query planning~\cite{oracle_rls}. Database engines differ in their ability to detect such cycles prior to
execution: most of the planners perform static cycle detection and raise an
error at policy-compilation time. 

\subsubsection*{Acyclic: Tiered Policy Generation}
It is imperative that the policy generation also allow a systematic way to prevent cyclic policy structure. 
We have the following rule to achieve that.

Let \(G_s\) be the schema dependency DAG, \(A(t)\) be the set of ancestor tables of table \(t\) in \(G_s\), and \(P_t\) be the policy selected for table \(t\). Let \(B\) denote the set of base tables. Policy generation is constrained by
$
\mathrm{Refs}(P_t) \subseteq A(t) \cup B.
$

The tier assignment is then defined recursively as
\[
\mathrm{tier}(t) =
\begin{cases}
0 & \text{if } t \in B, \\[4pt]
1 + \max_{u \in \mathrm{Refs}(P_t)} \mathrm{tier}(u) & \text{otherwise.}
\end{cases}
\]

Because every policy reference points only to tables in strictly lower tiers, the resulting policy-dependency graph is acyclic.

\section{Experimental Analysis}

\paragraph{Objective}
We instantiate our policy framework to generate practically meaningful, compositionally structured policies over the TPC-H benchmark, enforced through the standard analytical query workload using the implementation strategies discussed in Section~\ref{sect:pol_imple}. While the enforcement mechanisms are individually well-understood, its behavior under complex, nested policy compositions remains unstudied. Our goal is to close this gap by systematically characterizing the optimization behavior and performance limits of each strategy as policy complexity increases. Our code has been made publicly available\footnote{https://github.com/imtiyazuddin/secure26db.git
}.

\paragraph{System Setup:} Our setup includes 2 $\times$ Intel(R) Xeon(R) Silver 4114 CPUs @ 2.20GHz, 128 GB DDR4 RAM, and 8 TB HDD, running Ubuntu 24.04.4 LTS.  
We experimented with opensource {\bf PostgreSQL 18.2} and a commercial engine: Com-eng A \footnote{Name anonymized due to the "Dewitt" Clause}. We use the standard TPC-H benchmark at scale factor 1 for the underlying database.\footnote{We also conducted experiments with scale factor 10, but primarily report results at scale factor 1 due to the high runtime caused by query complexity induced via policies.} PostgreSQL requires explicit tuning, which we perform using the standard PGTune configuration. By default, experiments use standard primary-key and foreign-key indexes, given the OLAP focus of the user query workload.

\eat{
In most operational environments, the set of compliance policies is relatively stable and well-defined, whereas the workload of ad-hoc queries can vary significantly. Therefore, it is more effective to design indexes \emph{around the policy predicates} rather than around the (unknown) future query workload.
Accordingly, in addition to the usual primary-key (PK) and foreign-key (FK) indexes, we also consider creating indexes on the attributes referenced by policy predicates (i.e., the columns that appear in policy filters). These \emph{policy-coverage indexes} are intended to accelerate predicate evaluation during policy enforcement.}

\subsection{Policy Generation } 
We generate content-based policies over the TPC-H schema using our framework's policy model, grammar, and tiered acyclic composition structure. Policies are designed to carry genuine business semantics and realistic governance scenarios. We leverage the NL2SQL capabilities of GPT-5 by encoding the policy model, grammar rules, and tiered structure as structured prompts, guiding the model to produce syntactically valid and semantically meaningful SQL policy predicates. Each generated policy is validated against the TPC-H schema for correctness and the above mentioned constraints into the evaluation workload. We generate a suite of policies, the full list of which is available in the Git repository. A representative subset used in our evaluation is described below. \footnote{In this work, we limit one policy per base table due to limitation in PostgreSQL and other engines}

\subsubsection{Acyclic Policies}
Due to some engine limitation for handling cyclic policies, or ending up in infinite recursion, we generate the following non-cyclic policies:
\begin{itemize}[wide, labelwidth=!, labelindent=0pt]
    \item {\tt Customer:} Only customers who have high purchasing power, and only if they belong to the approved business regions. 

    \item {\tt Supplier:} Only suppliers with a good financial standing are included, and they must belong to the Europe region. 

    \item \texttt{Partsupp:} Only supplier-part inventory records are included when the available stock is plentiful, and the linked supplier is financially stable. 

    \item \texttt{Lineitem:} Only transaction line items are included when they are associated with high-priority orders and with part-supplier combinations that have plenty of stock. 
\end{itemize}

\subsubsection{Cyclic Policies}\label{sect:unres}
We also consider \emph{unrestricted} policies that permit cyclic inter-policy dependencies, which arise naturally in certain applications such as mutual visibility constraints or bidirectional access rules. While such cycles preclude the tiered acyclic structure assumed earlier, PostgreSQL's \texttt{SECURITY DEFINER} feature provides a safe enforcement path:  breaking the recursive policy application chain and preventing infinite evaluation loops.

\begin{itemize}[wide, labelwidth=!, labelindent=0pt]
    \item \texttt{Orders:} Only high-value orders are included. In business terms, an order is visible only if its total discounted sales value exceeds a defined revenue threshold. 

    \item \texttt{Customer:} Only customers who have participated in cross-country business within the same region are included. 

    \item \texttt{Part:} Only parts that have not been associated with unusually large individual order quantities are included. 

    \item \texttt{Supplier:} Only suppliers are included who are not linked to low-priced products below a defined price threshold. 
\end{itemize}

\eat{
These policies are activated in different combinations across the 22 TPC-H queries, based on the tables present in the {\tt FROM}-clause of each query. The query to policy mapping is listed in Table~\ref{tab:policy-to-tpch}. We mention which policy set, i.e., mapping, is used for which experiment in the respective details.

\begin{table}[t]

\centering
{\footnotesize
\begin{tabular}{@{}c l p{0.62\columnwidth}@{}}
\toprule
& \textbf{Policy-filtered table} & \textbf{TPC-H QID(s)} \\
\midrule
\multirow{4}{*}{\rotatebox{90}{Tiered}}
& customer & Q3, Q5, Q7, Q8, Q10, Q13, Q18, Q22 \\
& supplier & Q2, Q5, Q7--Q9, Q11, Q15, Q16, Q20, Q21 \\
& partsupp & Q2, Q9, Q11, Q16, Q20 \\
& lineitem & Q1, Q3--Q10, Q12, Q14--Q15, Q17--Q21 \\
\midrule
\multirow{4}{*}{\rotatebox{90}{Unrestricted}}
& orders   & Q3--Q5, Q7--Q10, Q12, Q13, Q18, Q21, Q22 \\
& customer & Q3, Q5, Q7, Q8, Q10, Q13, Q16, Q18, Q22 \\
& part     & Q2, Q8, Q9, Q14, Q16, Q17, Q19, Q20 \\
& supplier & Q2, Q5, Q7--Q9, Q11, Q15, Q16, Q20, Q21 \\
\bottomrule
\end{tabular}
}
\caption{Figure~\ref{fig:security_barrier_1g} - Policy-filtered Tables to QIDs}
\label{tab:policy-to-tpch}
\end{table}
}
\subsection{Policy Enforcement Methods: PostgreSQL}\label{sect:enforment_type}
As discussed in Section~\ref{sect:pol_imple}, we evaluate the following policy-enforcement strategies available in PostgreSQL:

\begin{enumerate}[wide, labelwidth=!, labelindent=0pt]
\item \textbf{Pure RLS / Indexed RLS:} Policies are enforced as Row-Level Security predicates, which PostgreSQL injects directly into query execution at the base-table level. To assess the benefit of planner-side optimization, we additionally evaluate \textbf{Indexed RLS}, where {\em policy-coverage indexes} are created to accelerate evaluation.

\item \textbf{Secure Views:} Since policies are typically stable and well-defined, each policy query is a view, and the original user query is rewritten to reference these views in place of the base tables.

\item \textbf{Query Rewrite:} Policy predicates are inlined directly into the user query, and the resulting combined query is rewritten using the state-of-the-art rewriter \texttt{LITHE}~\cite{lithe}, enabling cross-predicate optimization across the policy and query logic.

\item \textbf{UDF:} PostgreSQL User-Defined Functions (UDF) defined with \texttt{SECURITY DEFINER} treat the policy as an opaque black box, disabling cross-optimization between the policy and the user query. While transparent UDFs permit some cross-optimization, their performance is consistently inferior to query rewrite; we therefore include only the security definer variant in the evaluation.

Note that each of the above methods when referred to in the rest of the section is, by default, evaluated in a {\bf\em white-box setting}, which enables cross query–policy optimization. When this is not allowed, we denote it by {\bf\em black-box setting}. Our experiments do not use masking policies as they cannot be enforced using RLS.

\end{enumerate}

\subsection{Experiments \& Observations: PostgreSQL}

\subsubsection{\textbf{Cross Query-Policy Optimization: Enabled}}\label{sect:lithe}
\begin{figure}
    \centering
    \includegraphics[width=\linewidth]{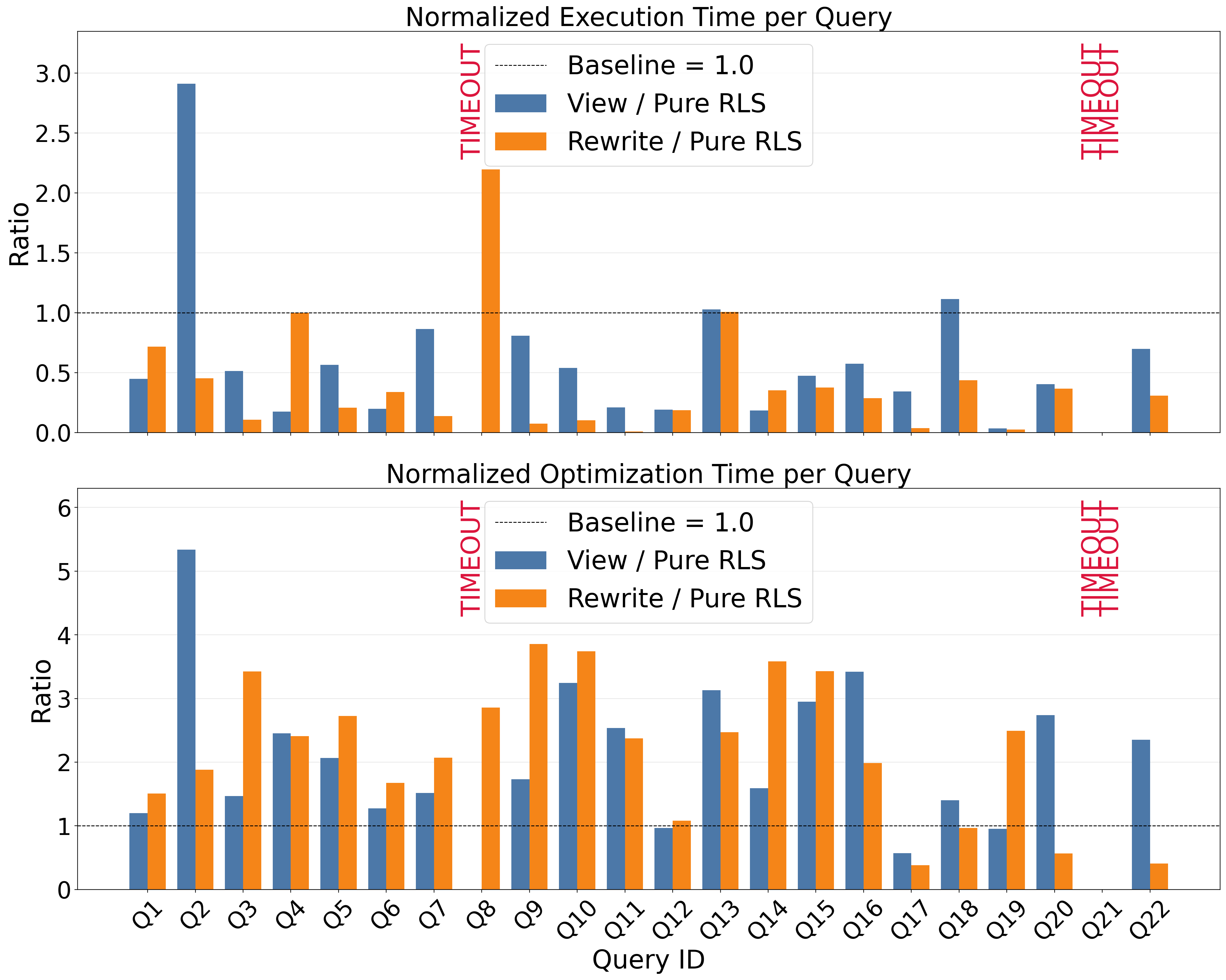}
    \caption{RLS vs. View vs. Query Rewrite for Acyclic Policies}
    \label{fig:lithe}
\end{figure}

We first evaluate enforcement strategies that treat policies as white-box, enabling cross-optimization between policy predicates and the user query. This includes Pure RLS,  Views, and Query Rewrite via \texttt{LITHE}. Since policies are structurally exposed to the optimizer, cyclic policies are excluded from this setting to avoid infinite recursion during predicate injection.
Figure~\ref{fig:lithe} reports normalized execution time (top) and optimization time (bottom) for Views and Query Rewrite, both normalized to Pure RLS as the baseline (ratio $< 1$ indicates faster than Pure RLS).

\paragraph{Execution time.}
Query Rewrite consistently outperforms Pure RLS across the majority of queries, confirming that explicit predicate inlining followed by algebraic simplification via \texttt{LITHE} allows the optimizer to make better join-ordering and filter-pushdown decisions than the implicit predicate injection used by RLS. The gains are most pronounced on join-heavy queries with correlated policy predicates—Q3, Q7, Q9, Q10, Q11, Q17, and Q19—where Pure RLS fails to exploit cross-predicate optimization opportunities. However, Query Rewrite times out on Q8 and Q21, indicating that aggressive inlining of deeply nested, tiered policies can produce query trees too large for the planner to handle within practical limits. In terms of overhead, \texttt{LITHE} uses a small number of LLM iterations to converge on the best rewritten query, which is a one-time cost per query–policy combination.

Overall,  Views also achieve significant execution improvements over Pure RLS on several queries—Q4, Q6, Q11, Q12, Q14, and Q19. The view cost is a one-time effort that is practically acceptable given that policies change infrequently. The fact that Query Rewrite outperforms Views on several queries reveals missed optimization opportunities by the planner.

\paragraph{Optimization time.}
Pure RLS spends the least planning time across most queries, as the injected predicates remain compact and do not expand the planner's search space. Both Views and Query Rewrite incur substantially higher planning overhead—often 2–4$\times$ that of Pure RLS—since view introduces additional scan nodes while query rewriting produces larger, more complex query trees that broaden the planner's enumeration space. This is the central tension of the white-box approach: better execution comes at the cost of higher planning overhead.

\paragraph{Effect of indexing.}
Indexed RLS yields only marginal improvement over Pure RLS. The acyclic policy predicates used here do not exhibit selectivities favorable for index use; moreover, acyclic policies are structurally simpler and largely lack the complexity that could benefit from index-based access paths.

\paragraph{Impact of Scale}
\begin{figure}
    \centering
    \includegraphics[width=\linewidth]{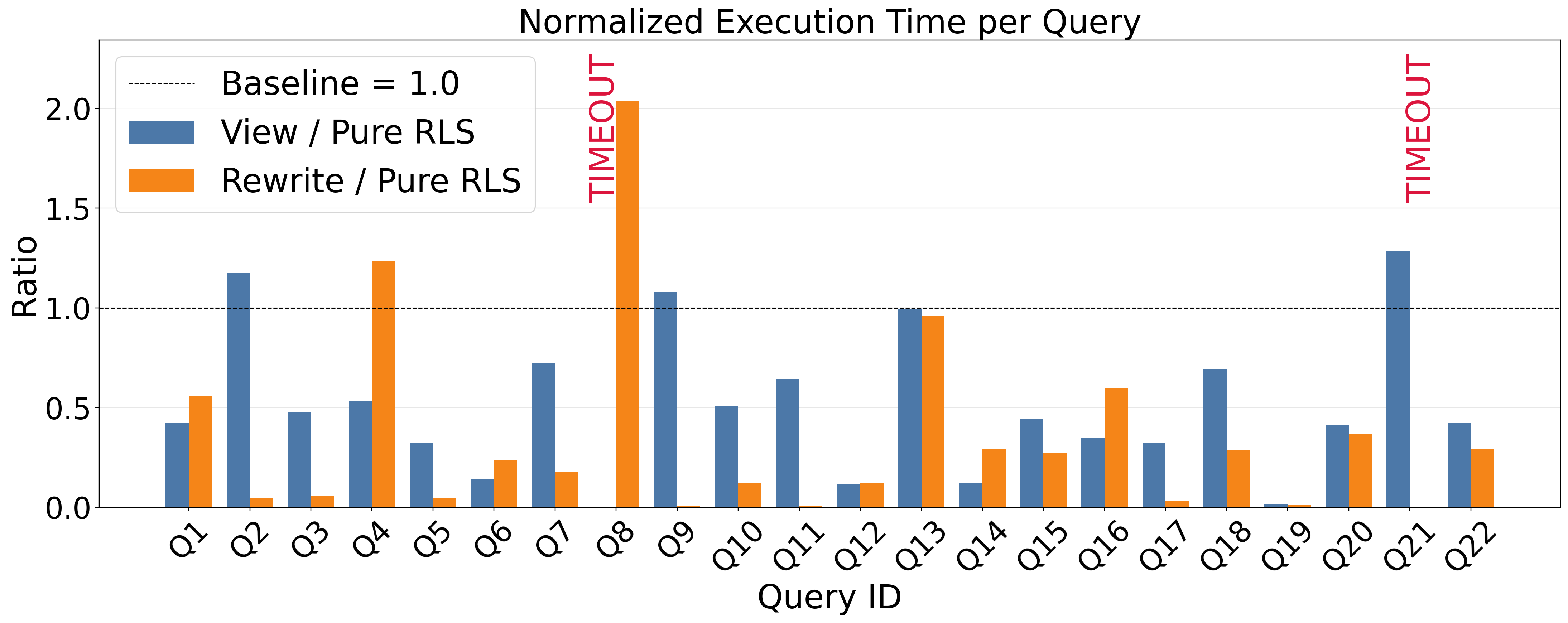}
    \caption{RLS vs. View vs. Rewrite on scale factor 10 data}
    \label{fig:10gb}
\end{figure}

To assess sensitivity to data volume, we repeat the evaluation at TPC-H scale factor 10 and report results in Figure~\ref{fig:10gb}. The relative ordering of the three strategies is preserved: both Views and Query Rewrite continue to outperform Pure RLS in execution time and robustness. Notably, the per-query performance profile remains largely stable across scale factors, suggesting that the observed behavior is driven by policy structure rather than data volume. There are, however, a few scale-dependent surprises: Q21 completes without timeout under Views at SF-10 (unlike at SF-1), and the View slowdown on Q2 narrows from approximately $3\times$ to under $1.5\times$. These exceptions suggest that at larger scales, increased cardinality may influence the optimizer plan better.

\paragraph{Takeaway.}
Query Rewrite via \texttt{LITHE} is the most effective enforcement strategy under the white-box setting, delivering consistently lower execution times at the cost of higher planning overhead and occasional planner timeouts on the most complex queries. Views offer a practical middle ground—stable and easier to deploy but leave missed optimization opportunities on the table. One can also think of using the above two techniques in conjunction for more performance benefits. Pure RLS, while the most predictable and planner-friendly, pays a systematic execution penalty as policy complexity grows. 
Finally, the relative performance profile of each strategy is largely independent of data volume, with only isolated queries showing scale-dependent behavior -- reinforcing that policy structure is the primary determinant of enforcement overhead.

\subsubsection{\textbf{Cross Query-Policy Optimization: Disabled}}

We now turn to the setting where cross-optimization between queries and policies is disabled. As discussed in Section~\ref{sect:pol_imple}, this is the appropriate choice when policies must be treated as black-box for security reasons. The opacity of the policy boundary prevents the engine from recursively expanding policy predicates, which also means PostgreSQL does not enter infinite recursion on cyclic policies. This setting therefore naturally accommodates unrestricted, cyclic policy structures, which are evaluated here. Note that in this scenario, UDF and pure RLS (black-box version) become synonymous.

\paragraph{Effect of indexing.}

\begin{figure}
    \centering
    \includegraphics[width=\linewidth]{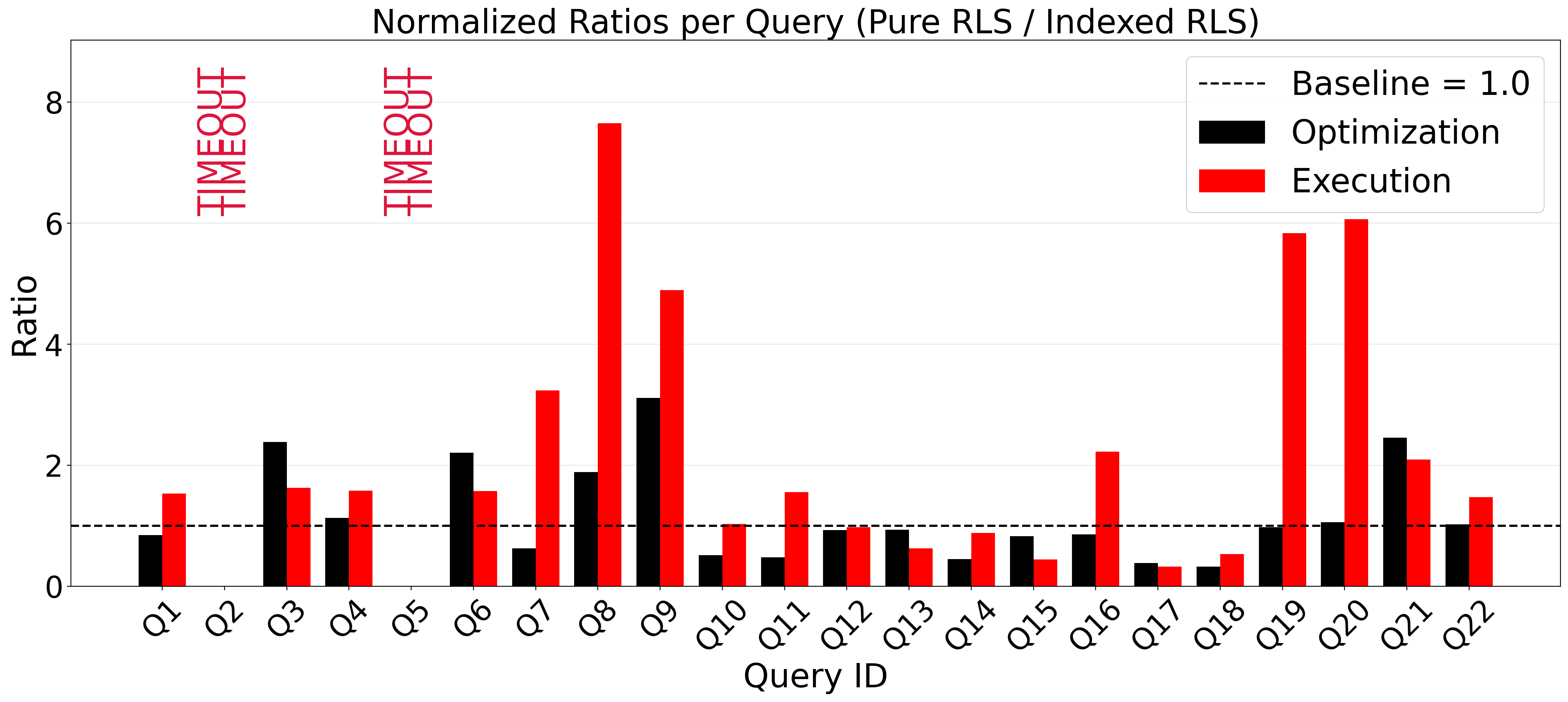}
    \caption{Impact of Policy-Aware Indexing on Black-box RLS}
    \label{fig:security_barrier_1g}
\end{figure}

In the acyclic setting, indexing offered negligible benefit. We now investigate whether cyclic policies—which are structurally more complex—create conditions where policy-aware indexing can provide meaningful speedup. In the Indexed RLS (black-box) configuration, indexes are created on all individual columns referenced by the policy queries. Figure~\ref{fig:security_barrier_1g} reports normalized execution and optimization time ratios of Pure RLS (black-box) with respect to Indexed RLS (black-box).

The most important result is the dominant overhead of Pure RLS (black-box) in \emph{execution}. This pattern is most pronounced on queries that activate multiple policies simultaneously. Q8, which triggers all four policies at once, exhibits one of the largest execution slowdowns in the workload. Similarly, Q5, Q7, and Q9, each combining three policies, are considerably more expensive under Pure RLS (black-box). Query Q20 combines two policies — also show elevated normalized execution times. Lastly, although Q19 has only one policy, it uses a universal policy predicate, which evaluates a correlated predicate and then negates it, leading to high runtime. This policy also contributes to the slowdown of Q20.
Taken together, these results establish that policy composition is a primary driver of execution overhead and that indexing provides meaningful relief as compositional complexity grows. 

Importantly, Pure RLS (black-box) times out on Q2 and Q5, while Indexed RLS (black-box) completes both successfully. This demonstrates that the benefit of policy-aware indexing extends beyond average latency reduction—it also improves robustness and completion reliability on the most demanding queries. That said, Indexed RLS (black-box) does not uniformly dominate: Q13, Q15, Q17, and Q18 all execute faster under Pure RLS (black-box), reflecting query-specific plan choices where additional indexes do not align with the preferred access patterns. Nevertheless, across the full workload, Indexed RLS (black-box) offers a markedly better worst-case profile and more stable completion behavior.

Optimization-time behavior is more mixed: the additional indexes do not consistently increase or decrease planning overhead, suggesting that index availability alone does not systematically perturb the planner's search process.
 




\paragraph{Black-box Pure RLS vs.  Views: Unrestricted Policies}
\begin{figure}
    \centering
    \includegraphics[width=\linewidth]{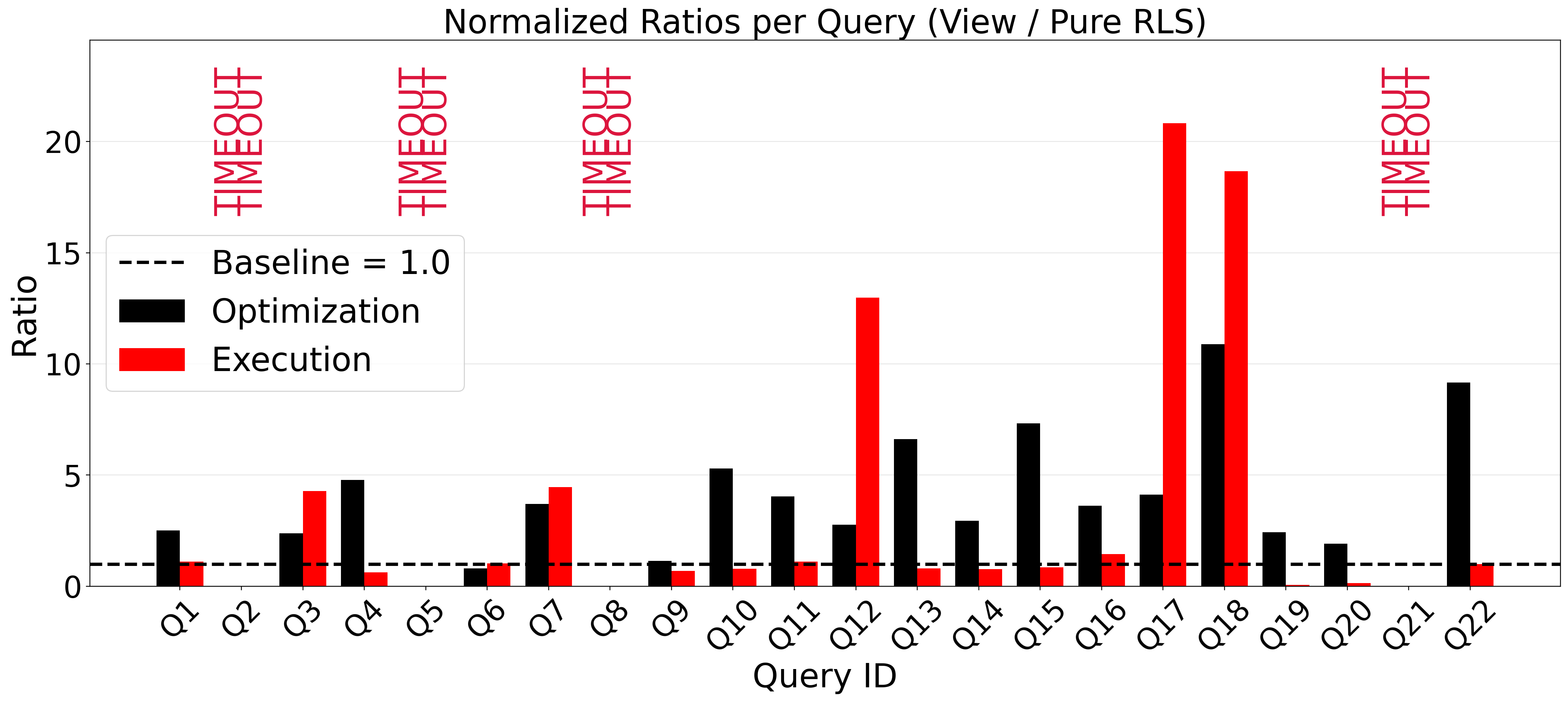}
    \caption{Black-box RLS vs. View for Cyclic Policy}
    \label{fig:rls_vs_view}
\end{figure}

Figure~\ref{fig:rls_vs_view} compares Pure RLS (black-box) and Security Barrier Views across planning and execution time for the unrestricted (cyclic) policy set.  To handle cycles safely, RLS predicates are wrapped in \texttt{SECURITY DEFINER} functions~\cite{postgres_create_function_security_definer}, and views as security barrier views~\cite{postgresql_create_view_security_barrier_2026}, which impose the same kind of optimization boundary. Since \texttt{LITHE} does not respect such boundaries, query rewrite is excluded from this experiment.
Pure RLS (black-box) provides more stable
trade-off across the workload, while views introduce consistently
higher planning overhead and more variable execution behavior. 

\begin{table}[h]

\begin{tabular}{@{}c@{}}
\begin{tabular}{|c|c|}
\hline
\textbf{$p(o)$} & \textbf{$\neg p(o)$} \\[4pt]
\hline
{\footnotesize
\(
\begin{aligned}
(o\_orderstatus = \texttt{'F'}) \land \\
(o\_totalprice > 100000) \land \\
(o\_orderdate \ge \texttt{1995-01-01})
\end{aligned}
\)
}
&
{\small
\(
\begin{aligned}
(o\_orderstatus \neq \texttt{'F'}) \land \\
(o\_totalprice \le 100000) \land \\
(o\_orderdate < \texttt{1995-01-01})
\end{aligned}
\)
}
\\
\hline
\end{tabular}
\\
{\small
\begin{tabular}{@{}lrr@{}}
\toprule
Mechanism & Access path & Slowdown \\
\midrule
Predicate Injection     & Parallel semi-joins           & $1.00\times$ \\
Security Definer Policy & Index-only scan + UDF filter & $55.09\times$ \\
\bottomrule
\end{tabular}
}
\end{tabular}
\caption{Case with $p(o) \wedge \neg p(o)$ Policy}
\label{tab:plan123}
\end{table}

\paragraph{Takeaway.}
Under black-box setting, policy-aware indexing significantly improves both robustness and worst-case execution latency for cyclic policies, while Pure RLS (black-box) remains more stable and planning-efficient than Security Barrier Views. Policy composition depth is the dominant cost driver in this setting. Future optimizers should use cost-based optimization to select between view-based and pure RLS approaches for a given query.

\noindent
\begin{tabular}{|l|l|}
\hline
\textbf{$p(o)$} & \textbf{$q(o)$} \\[4pt]
\hline
{\footnotesize
\(
\begin{aligned}
(o\_totalprice > 150000) \land \\
(o\_orderstatus = \texttt{'F'}) \land \\
(o\_orderdate > \texttt{1994-12-31}) \land \\
\exists o_2 \,(o_2.o\_custkey = o.o\_custkey \\
\qquad \land\; o_2.o\_orderstatus = \texttt{'F'})
\end{aligned}
\)
}
&
{\footnotesize
\(
\begin{aligned}
(o\_totalprice > 50000) \land \\
(o\_orderdate > \texttt{1993-12-31}) \land \\
\exists o_2 \,(o_2.o\_custkey = o.o\_custkey \\
\qquad \land\; o_2.o\_orderstatus = \texttt{'F'})
\end{aligned}
\)
}
\\
\hline
\end{tabular}
\captionof{table}{Atomic Policies for Composition}
\label{tab:p_n_q}

\par\noindent
{\small
\begin{tabular}{@{}lcc@{}}
\toprule
Predicate Composition & Security Definer & Predicate Injection \\
\midrule
$p \land \neg q$ & $1.00\times$ & $1.00\times$ \\
$\neg q \land p$ & $2.74\times$ & $1.83\times$ \\
\bottomrule
\end{tabular}
}
\captionof{table}{Slowdown caused by predicate order}
\label{tab:pred-order}

\par\noindent
{\small
\begin{tabular}{@{}p{1cm} p{1.2cm} p{4cm} r@{}}
\toprule
Order & Mechanism & Access path & Slowdown \\
\midrule
$p \wedge \neg q$ &
Security Definer Policy &
Parallel index-only scan using \texttt{o\_orderstatus} index, partial/final aggregate &
$1.00\times$ \\

$p \wedge \neg q$ &
Predicate Injection &
Bitmap heap scan with \texttt{BitmapAnd} over all indexes, + {\bf one} subplan &
$8.62\times$ \\

\midrule

$\neg q \wedge p$ &
Security Definer Policy &
Parallel index-only scan using \texttt{o\_orderstatus} index, partial/final aggregate &
$1.00\times$ \\

$\neg q \wedge p$ &
Predicate Injection &
Bitmap heap scan with \texttt{BitmapAnd} over all indexes, + {\bf two} subplans &
$5.75\times$ \\
\bottomrule
\end{tabular}
}
\captionof{table}{Impact of Policy-Coverage Indexing}
\label{tab:secdef-injection-combined}

\subsubsection{\textbf{Stress-testing Optimizers with Policy Complexity}}
In this category of experiments, we test how the RLS mechanism handles the simplification of logical connections among policy predicates. We take the fastest implementation as 1x and compare the relative slowdown of the other.

First, we check the logical AND-ing of contradictory policies. The predicates are on \texttt{orders} tables, as in Table~\ref{tab:plan123}.

We have a composite policy $p(o) \wedge \neg p(o)$, where each policy is either implemented using (i) security definer function to be used as RLS, or (ii) manual predicates injected with the user query -- equivalent to standard RLS. Both versions use only PK+FK indexes.

The main observation is listed in Table~\ref{tab:plan123}.
RLS enabled with black-box functions hides predicate semantics inside scalar functions. As a result, PostgreSQL cannot rewrite the logic into joins; it performs an index-only scan over all 1.5M rows and applies the opaque function filter tuple-by-tuple. This yields the worst behavior despite zero heap fetches. On the other hand, when the predicates are visible in the query plan, the planner converts the same test into set-oriented parallel semi-joins, so the engine evaluates $p$ and $q$ once as relations and combines them efficiently. Therefore, 
predicate visibility and set-based rewriting matter far more than nominal index use.

Next, we study a composite policy that involves negations. The component predicates are listed above, In the composition, we test {\tt EXISTS} $p(o)$ $\wedge$ {\tt NOT EXISTS} $q(o)$ where both orders of the predicates are tested for each mechanism. 
{\em We also used policy-coverage indexes here}. The results are listed in Table~\ref{tab:pred-order}.
The key observation is that order of the {\tt NOT} predicate causes severe execution slowdown, in both the mechanicsim. In predicate injection, there was no change in the plan shape. It is consistent with executor-level variance (e.g., \texttt{Gather}/parallel coordination) rather than a true optimization effect. 
Because $q$ is subsumed by $p$, $\neg q$ was not executed, hence, obtaining faster result.
On the other hand, black box functions could not have such optimization, thereby gaining higher slowdown.

One more interesting point here is the effect of indexing. Predicate injection is substantially slower than the security-definer formulation for both predicate orders, as listed in Table~\ref{tab:secdef-injection-combined}. The security-definer/RLS plan uses a parallel index-only scan. In contrast, injected predicates used a bitmap heap scan and large subplans. 

\eat{
\begin{table}[t]
\centering
{\small
\begin{tabular}{lcc}
\toprule
Predicate Composition & Security Definer & Predicate Injection \\
\midrule
$p \land \neg q$ & $1.00\times$ & $1.00\times$ \\
$\neg q \land p$ & $2.74\times$ & $1.83\times$ \\
\bottomrule
\end{tabular}
}
\caption{Slowdown caused by predicate order}
\label{tab:pred-order}
\end{table}

\begin{table}[t]
\centering
{\small
\begin{tabular}{p{1cm} p{1.2cm} p{4.2cm} r}
\toprule
Order & Mechanism & Access path & Slowdown \\
\midrule
$p \wedge \neg q$ &
Security Definer Policy &
Parallel index-only scan on \texttt{idx\_orders\_orderstatus}, partial/final aggregate&
$1.00\times$ \\

$p \wedge \neg q$ &
Predicate Injection &
Bitmap heap scan with \texttt{BitmapAnd} over all indexes, + {\bf one} hashed subplan &
$8.62\times$ \\

\midrule

$\neg q \wedge p$ &
Security Definer Policy &
Parallel index-only scan on \texttt{idx\_orders\_orderstatus}, partial/final aggregate&
$1.00\times$ \\

$\neg q \wedge p$ &
Predicate Injection &
Bitmap heap scan with \texttt{BitmapAnd} over all indexes, + {\bf two} hashed subplans &
$5.75\times$ \\
\bottomrule
\end{tabular}
}
\caption{Impact of Policy-Coverage Indexing}
\label{tab:secdef-injection-combined}
\end{table}
}

\subsection{Commercial Database Engine}
\begin{figure}
    \centering
    \begin{tabular}{c c}
       \includegraphics[width=0.4\linewidth]{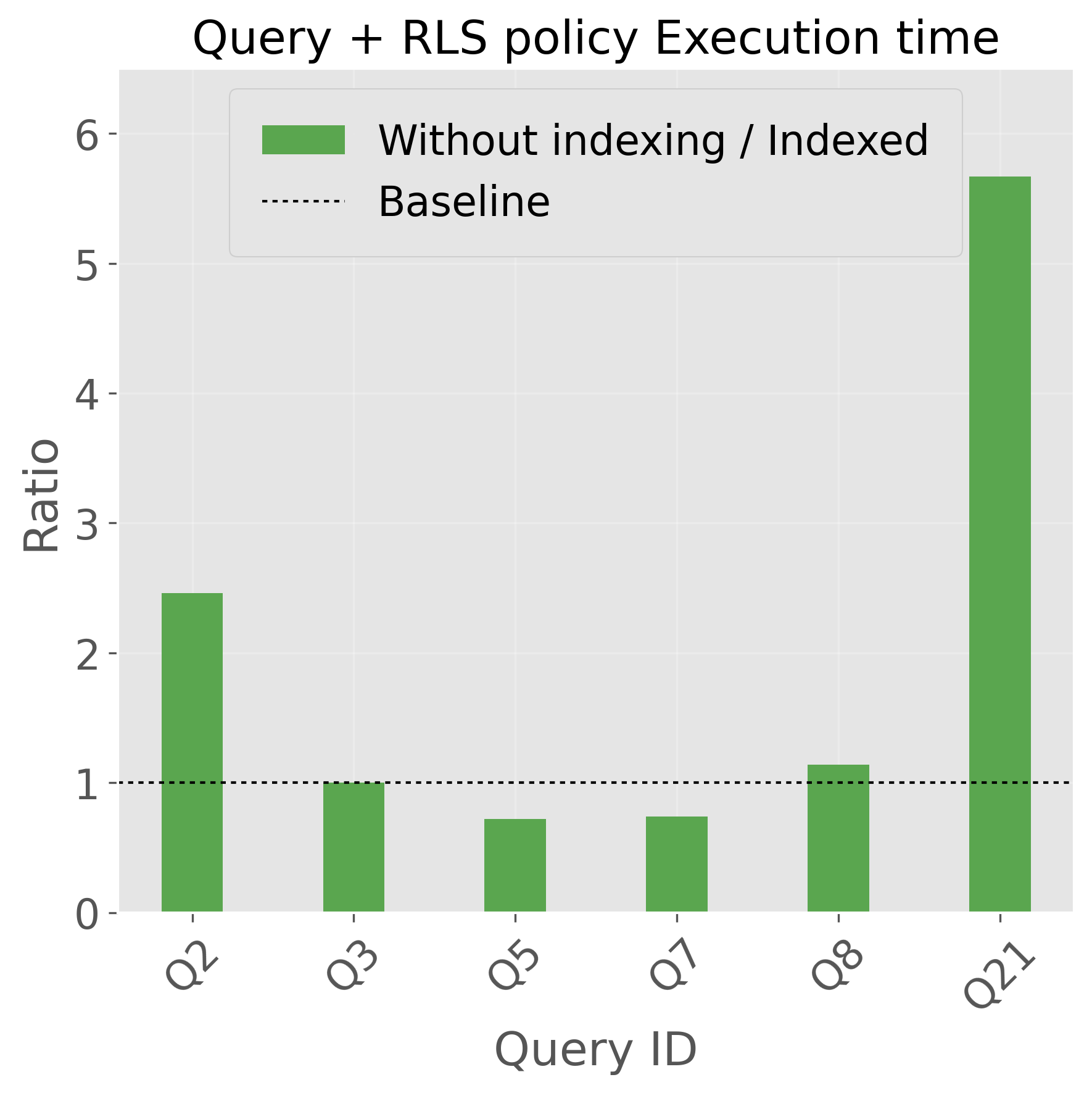}  &  \includegraphics[width=0.4\linewidth]{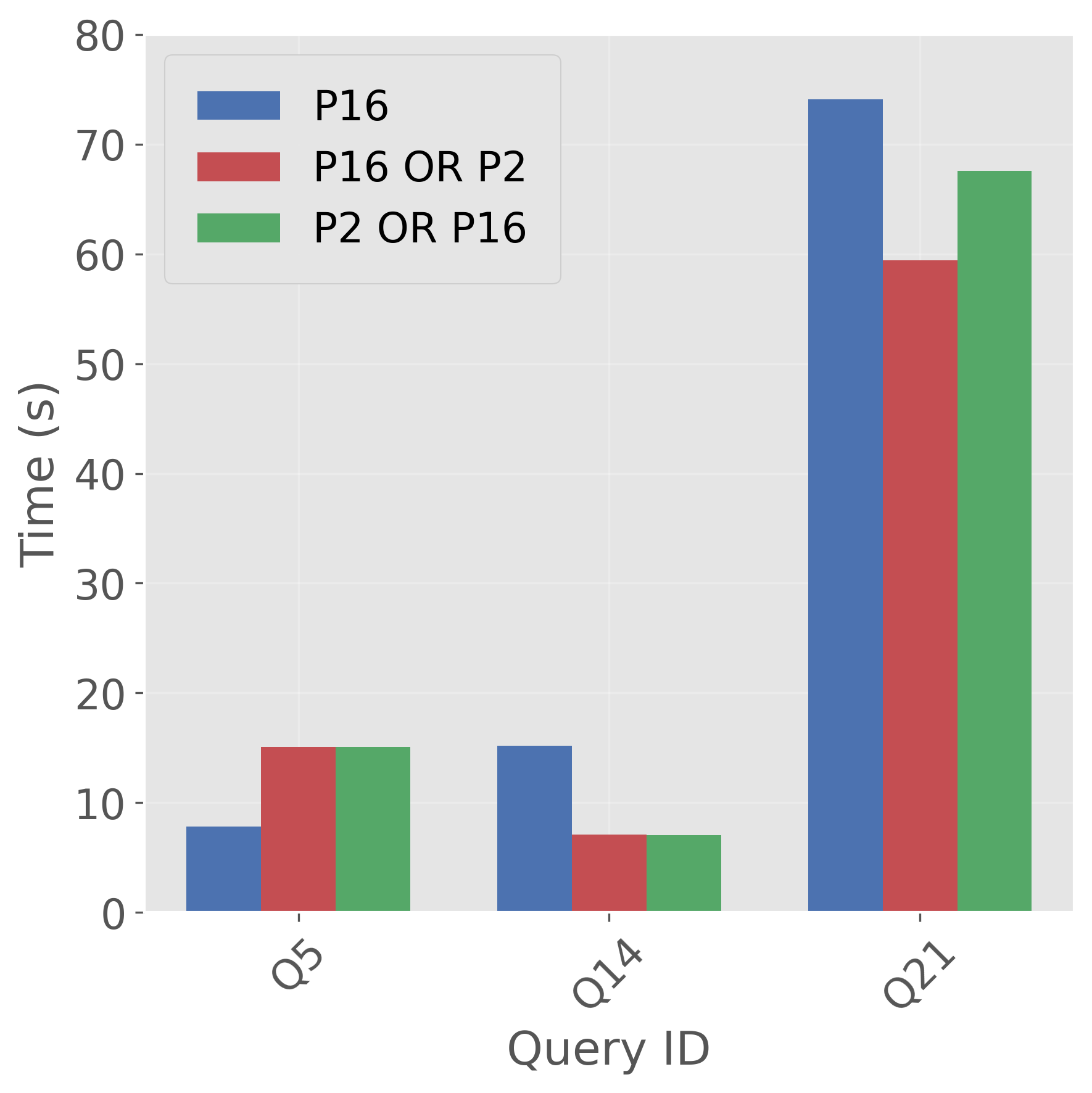}\\
      (a) Pure RLS/Indexed RLS   & (b) Logical {\tt OR} using RLS\\
    \end{tabular}
    \caption{RLS Policy performance in Comm-Engine A}
        \label{fig:vpd}
\end{figure}

We also looked at a Com-Engine A, specifically on the impact of index and order of predicates in logical combination in a feature similar to PostgreSQL RLS. Similar to PostgreSQL, the injected predicates also effectively use the indexes, as can be observed in Figure~\ref{fig:vpd}(a). Moreover, we investigate the effect of predicate ordering in the logical {\tt OR} combination, using one simple and one expensive predicate -- labeled as P2 and P16, respectively, in Figure~\ref{fig:vpd}(b). It is worth noting that the combined policy incurs both faster and slower execution times, depending on the query. Also, the order of the predicate matters in most of the cases. 

\section{Related Work}

Prior work on database compliance spans policy-aware query processing, query-time compliance checking, disclosure control, and policy specification. Under our framework, GDPRbench~\cite{saeed2019gdprbench}, and XACML~\cite{oasis_xacml} primarily capture \emph{attribute-predicate} policies from our atomic policy structures: GDPRbench models GDPR rights such as consent, erasure, portability, and purpose limitation as SQL workloads;  XACML provides a general declarative ABAC language. Beedkar et al.~\cite{beedkar2021compliant} and Schwab et al.~\cite{schwab2021policy} use both \emph{attribute} and \emph{grouping/aggregate} predicates: the former introduce compliant geo-distributed query processing, in which SQL-dataflow policies are incorporated into query planning as location constraints, while the latter checks query compliance against policy rules using similarity-based matching. Mascara~\cite{poepsellemaitre2024disclosure} enforces data disclosure policies by rewriting SPJG SQL queries and applying intuitive {\em data masks} for context-aware partial disclosure of sensitive attributes based on user roles. The {\em transformation enforcement action} -- $\mu$ functions that do to map to $\bot$ -- captures the same semantics of data-masking. 

Finally, AND/OR composition of predicates is also observed in Sieve~\cite{pappachan2020sieve}, which scales fine-grained enforcement for large GDPR/CCPA rule sets through query-context filtering. However, Sieve derives its policy rules for IoT use cases, where the rules are predominantly simple value-based predicates. Composition of security policies is approached in the security domain, such as in SePL~\cite{mejri2017formal}. The composition framework is a modal logic-based language that allows union (least restrictive), intersection (most restrictive), prioritized override, sequential chaining, and conditional activation of diverse actions (read/write, execute, connect, delegate, encrypt/decrypt)s. Our framework brings a similar capability of combining policy queries in the database context.
Similar contrast can be found in other security policy frameworks, such as in ~\cite{epal_wiki} and ~\cite{damianou2001ponder}.

\section{Conclusions \& Future Work}

We presented a framework for generating and reasoning about structurally complex content-based compliance policies, through a formal policy model and expressive grammar that supports both \emph{white-box} and \emph{black-box} enforcement methods. Using this framework, we augmented TPC-H with structured policy workloads and evaluated four PostgreSQL enforcement strategies. Our central finding is that policy structure is the primary driver of enforcement overhead, and that no single strategy dominates across all settings. These results expose a fundamental gap in current optimizer design.

Based on our findings, we identify several directions for future work and database engine design. On the system side, actionable improvements include: \emph{(i)} policy-aware cardinality estimation, where the planner models policy predicate selectivity jointly with query predicates; \emph{(ii)} a policy-aware index advisor that recommends indexes on policy-referenced columns as a first-class optimization target; \emph{(iii)} partial predicate pushdown through security barriers, relaxing the current all-or-nothing barrier semantics for well-typed, non-sensitive predicates; and \emph{(iv)}  enabling the engine to natively select black-box enforcement for cyclic RLS-policies without requiring manual definitions of black-box wrappers. 

On the formalism side, our current model assumes atomic policies act on disjoint tuple sets; relaxing this to support overlapping scopes will require priority or conflict-resolution rules. Finally, query-rewriting-aligned techniques that translate compliance policies into regulation-aware transformations while preserving execution robustness remain an important open problem.

\section*{Acknowledgment}
This work was partially funded by the Privacy-Preserving Data Processing and Exchange for Sensitive Data in the National Digital Public Infrastructure (P3DX) project of Srinivas Vivek.

\section*{Declaration of using AI}

We used popular GPT models for coding and writing assistance but assume full responsibility for the accuracy of the work.

\bibliographystyle{abbrv}
\bibliography{ref}

\appendix

\section*{Appendix}\label{Sect:appendix}
\subsection*{Examples of Atomic Policy Predicates wrt Section~\ref{sect:atomic}}
\begin{enumerate}[wide, labelwidth=!, labelindent=0pt]
    \item {\bf Attribute Predicate:} In \texttt{TPC-H} setting, a customer may be allowed to see only those lineitems that belong to her orders and are supplied by suppliers from the same nation as the customer. Therefore, we have $R$ = {\tt lineitem}, and $\pi$ is the identity projection over all attributes of {\tt lineitem}. Using $l$ as its alias, the policy query corresponding to the authorized tuples of {\tt lineitem} is given as follows:
\begin{quote}\footnotesize
\begin{verbatim}
EXISTS (SELECT 1
    FROM orders o, customer c, supplier s
    WHERE l.l_orderkey = o.o_orderkey
      AND o.o_custkey = c.c_custkey
      AND l.l_suppkey = s.s_suppkey
      AND c.c_name = :current_user
      AND c.c_nationkey = s.s_nationkey)
\end{verbatim}
\item {\bf Existential Predicate:} Any query may be allowed to fetch data of only those customers who have placed at least one high-value order. Therefore, we have $R$ = {\tt customer}, and $\pi$ is the identity projection over all attributes of {\tt customer}. Using $c$ as its alias, $Q(c)$ is given as follows:
\begin{quote}\footnotesize
\begin{verbatim}
EXISTS (SELECT 1 FROM orders o
    WHERE o.o_custkey = c.c_custkey AND o.o_totalprice > 1000)
\end{verbatim}
\end{quote}
\item {\bf Universal Predicate:} Any query may be allowed to fetch data of only those customers who have not returned any ordered item. Therefore, we have $R$ = {\tt customer}, and $\pi$ is the identity projection over all attributes of {\tt customer}. Using $c$ as its alias, $Q(c)$ is given as follows:

\begin{quote}\footnotesize
\begin{verbatim}
NOT EXISTS (SELECT 1
    FROM orders o, lineitem l
    WHERE o.o_custkey = c.c_custkey
      AND l.l_orderkey = o.o_orderkey
      AND l.l_returnflag = 'R')
\end{verbatim}
\end{quote}
\item {\bf Grouping/Aggregate Predicate:} Any query may be allowed to fetch data of only those customers who have purchased items from at least three distinct suppliers. Therefore, we have $R$ = {\tt customer}, and $\pi$ is the identity projection over all attributes of {\tt customer}. Using $c$ as its alias, $Q(c)$ is given as follows:

\begin{quote}\footnotesize
\begin{verbatim}
EXISTS (SELECT 1
    FROM orders o, lineitem l
    WHERE o.o_custkey = c.c_custkey
      AND l.l_orderkey = o.o_orderkey
    GROUP BY o.o_custkey
    HAVING COUNT(DISTINCT l.l_suppkey) >= 3)
\end{verbatim}
\end{quote}
\item {\bf Statistical Predicate:} Queries may be allowed to use data of only those {\tt lineitem} tuples whose discounted revenue contributes at least 20\% of the total revenue of their corresponding order. Therefore, we have $R$ = {\tt lineitem}, and $\pi$ is the identity projection over all attributes of {\tt lineitem}. Using $l$ as its alias, $Q(l)$ is given as follows:
\begin{quote}\footnotesize
\begin{verbatim}
l.l_extendedprice * (1 - l.l_discount) >= 0.2 * (
    SELECT SUM(l2.l_extendedprice * (1 - l2.l_discount))
    FROM lineitem l2
    WHERE l2.l_orderkey = l.l_orderkey)
\end{verbatim}
\end{quote}
\end{quote}

\end{enumerate}
\subsection*{SQL for the Composite Policy in Section~\ref{sect:example}}
Here, {\tt customer\_status} is the relation obtained by evaluating $\mathcal{P}$. 
CTEs {\tt customer\_spending} captures the evaluation of the LHS of the policy predicate $Q_1$, and {\tt customer\_returns} captures $Q_2$. The three disjoint selections $\sigma_{Q_1 \wedge \neg Q_2}(R^D)$, $
\sigma_{Q_1 \wedge Q_2}(R^D)$, and $
\sigma_{\neg Q_1}(R^D)
$ are present inside the body of this CTE. Finally, the projections $\pi$, $\mu_{\text{partial}}$, and $\mu_{\text{NULL}}$ correspond to the {\tt SELECT}-clause applied on {\tt customer\_status}, in the final part of the query.

\begin{quote}
\footnotesize
\begin{verbatim}
WITH customer_spending AS (
    SELECT
        c.*,
        COALESCE(SUM(l.l_extendedprice * (1 - l.l_discount)), 0) 
        AS total_spent
    FROM
        customer c, orders o, lineitem l
    WHERE
        c.c_custkey = o.o_custkey
        AND o.o_orderkey = l.l_orderkey
    GROUP BY
        c.c_custkey, c.c_name, c.c_address,
        c.c_nationkey, c.c_phone, c.c_acctbal,
        c.c_mktsegment, c.c_comment
),

customer_returns AS (
    SELECT DISTINCT
        o.o_custkey
    FROM
        orders o,
        lineitem l
    WHERE
        o.o_orderkey = l.l_orderkey
        AND l.l_returnflag = 'R'
),

customer_status AS (
    SELECT
        cs.*,
        CASE
            WHEN cs.total_spent > 5000
                 AND cr.o_custkey IS NULL
                THEN 'FULL'

            WHEN cs.total_spent > 5000
                 AND cr.o_custkey IS NOT NULL
                THEN 'PARTIAL'

            ELSE 'NONE'
        END AS visibility_level
    FROM
        customer_spending cs
        LEFT JOIN customer_returns cr
            ON cs.c_custkey = cr.o_custkey
)

SELECT
    c_custkey,
    c_name,

    CASE
        WHEN visibility_level = 'FULL'
            THEN c_address
        WHEN visibility_level = 'PARTIAL'
            THEN 'MASKED'
        ELSE NULL
    END AS c_address,

    CASE
        WHEN visibility_level = 'FULL'
            THEN c_phone
        WHEN visibility_level = 'PARTIAL'
            THEN 'MASKED'
        ELSE NULL
    END AS c_phone,

    CASE
        WHEN visibility_level IN ('FULL', 'PARTIAL')
            THEN c_nationkey
        ELSE NULL
    END AS c_nationkey,

    CASE
        WHEN visibility_level IN ('FULL', 'PARTIAL')
            THEN c_acctbal
        ELSE NULL
    END AS c_acctbal,

    CASE
        WHEN visibility_level IN ('FULL', 'PARTIAL')
            THEN c_mktsegment
        ELSE NULL
    END AS c_mktsegment,

    CASE
        WHEN visibility_level IN ('FULL', 'PARTIAL')
            THEN c_comment
        ELSE NULL
    END AS c_comment

FROM customer_status;
\end{verbatim}
\end{quote}

\end{document}